\documentclass[a4paper]{PoS}

\usepackage[utf8]{inputenc}
\usepackage{amsmath,amssymb,amsfonts}
\usepackage{bm}
\usepackage{array}
\usepackage{caption}
\usepackage{wrapfig}

\title{Phenomenology using PARTONS software of Generalized Parton Distribution models built from Light-Front Wave-functions}

\ShortTitle{Phenomenology of GPD models built from LFWFs}

\author{\speaker{Nabil Chouika}\thanks{The speaker is grateful to C.~Lorcé for fruitful exchanges and valuable inputs, and would like to thank his collaborators C.~Mezrag, H.~Moutarde, J.~Rodr\'iguez-Quintero and P.~Sznajder.
                                       This work is partly supported by he GDR QCD “Chromodynamique Quantique”, the ANR-12-MONU-0008-01 “PARTONS” and the Commissariat à l’Énergie Atomique et aux Énergies Alternatives.}\\
        IRFU, CEA, Universit\'e Paris-Saclay, F-91191 Gif-sur-Yvette, France\\
        E-mail: \email{nabil.chouika@cea.fr}}

\abstract{We present a procedure aiming at extending to the full kinematic domain a Generalized Parton Distribution obtained from a finite order truncation in Fock space.
This method allows to fulfill both polynomiality and positivity at the same time and can be applied to any given models of Light-front wave-functions.
In particular, we illustrate this on a three-body truncated wave-function of the chiral quark soliton model and show how a systematic phenomenology of GPD models based on LFWFs can be achieved with the help of the PARTONS framework, here using DVCS data.
This paves the way for a unified phenomenology of GPDs and TMDs at the level of LFWFs, with the final goal of hadron tomography.}

\FullConference{XXVI International Workshop on Deep-Inelastic Scattering and Related Subjects (DIS2018)\\
        16-20 April 2018\\
        Kobe, Japan}

\newcommand{\vperp}[2][]{\ensuremath{\bm{#2}_{\perp #1}}}
\newcommand{\drm}[0]{\mathrm{d}}

\begin{document}

\section{Introduction}

Generalized Parton Distributions (GPDs)~\cite{Mueller:1998fv, Radyushkin:1996nd, Ji:1996nm} encode the correlations between longitudinal momentum and transverse position of partons inside hadrons and can give access to a picture of the nucleon structure in 2+1 dimensions.
They have been studied theoretically and experimentally for almost two decades and a new experimental era is starting (at JLab and COMPASS currently, and in the future at an EIC) to extract them.

We can remind their definition in terms of a non-diagonal matrix element of a bi-local operator, in the simple case of a chiral-even twist-2 quark GPD of the pion:
\begin{equation}
  \label{eq:pion-quark-gpd-def}
  H^{q}\left(x,\xi,t\right)= \frac{1}{2}\int\frac{\drm z^{-}}{2\,\pi}\,e^{i\,x\,P^{+}z^{-}} 
                             \left. \left\langle \pi, P + \frac{\Delta}{2} \right|
                             \bar{\psi}^q\left(-\frac z 2\right) \gamma^{+} \psi^q\left(\frac z 2\right) \left| \pi, P-\frac{\Delta}{2}\right\rangle
                             \right|_{\substack{z^{+}=0\\z_{\perp}=0}} \, ,
\end{equation}
where $t = \Delta^2$ is the Mandelstam variable of momentum transfer and $\xi=-\frac{\Delta^{+}}{2\,P^{+}}$ is the skewness variable.
In the case of the nucleon, other chiral-even GPDs are of interest: $E$ (nucleon helicity flip), $\tilde{H}$ and $\tilde{E}$ (polarized GPDs).
We will mostly keep the simple pion case afterwards.

One of the main incentives to study GPDs is the probabilistic interpretation of their limit of zero skewness $H(x,\xi=0,t)$ \cite{Burkardt:2000za}, related by Fourier transform to a number density of partons
\begin{equation}
  \rho^q\left(x,\vperp{b}\right) =
  \int\frac{\mathrm{d}^{2}\vperp{\Delta}}{\left(2\pi\right)^{2}}\,e^{-i\,\vperp{b}\cdot\vperp{\Delta}}\,
  H^{q}\left(x,0,-\vperp{\Delta}^{2}\right) \,
  \label{eq:prob_density}
\end{equation}
of longitudinal momentum fraction $x$ and transverse position $\vperp{b}$.
Unfortunately, we currently have only an indirect access to these objects through convolutions of the form~\cite{Belitsky:2001ns}
\begin{equation}
  \label{eq:CFF-H-convolution}
  \mathcal{H}\left(\xi, t, Q^2\right) = \int_{-1}^1 \frac{\drm x}{\xi} \, \sum_{a=g,u,d,\ldots}
  C^a\left(\frac{x}{\xi},\frac{Q^2}{\mu_F^2}, \alpha_S\left(\mu_F^2\right)\right) H^a\left(x,\xi,t,\mu_F^2\right) \, .
\end{equation}
where $\mathcal{H}$ is a Compton Form Factor (CFF) associated to the GPD $H$, $C$ is a hard scattering kernel calculated at a given order in perturbation theory, $\mu_F$ is the factorization scale , $\alpha_S$ is the strong running coupling and $Q^2$ is the virtuality of the photon probing the hadron for instance in a process such as Deeply Virtual Compton Scattering (DVCS).
Moreover, only finite values of $\xi \in \left[\xi_{\mathrm{min}},\xi_{\mathrm{max}}\right]$ are accessible.
An extrapolation to vanishing skewness is therefore necessary on top of the already difficult access to the $x$-dependence.

Consequently, one of the main theoretical challenges in the field is to produce models of GPDs that are both polynomial and positive\footnote{For the sake of shortness, we will not go into the details of the other properties and constraints and we will consider only these two important ones. See \emph{e.g.} the reviews~\cite{Diehl:2003ny,Belitsky:2005qn} for details.}.
These constraints are important to be able to extrapolate accurately the information given by experimental data and can be stated as follows:
\begin{description}
\item[Polynomiality] The Mellin moments $\int_{-1}^{1}\mathrm{d}x\,x^{m}\,H\left(x,\xi,t\right)$ of a GPD $H$ are polynomials in the skewness variable $\xi$, of degree at most $m+1$.
This is related to Lorentz covariance.
\item[Positivity] The GPD is bounded by inequalities of the form~\cite{Pire:1998nw, Diehl:2003ny}
\begin{equation}
  \label{eq:positivity}
  \left|H^q(x,\xi,t)\right|_{x \ge \xi} \le
  \sqrt{q\left(\frac{x-\xi}{1-\xi}\right)q\left(\frac{x+\xi}{1+\xi}\right)} \, ,
\end{equation}
where the bound of the quark GPD $H^q$ is given in terms of its forward limit, \emph{i.e.}~the PDF $q$.
This reflects the positivity of a norm in a Hilbert space, where a Cauchy-Schwarz theorem can be applied.
\end{description}

There are two main roads to take in a quantum field theoretical framework leading to a computation of GPDs.
The first one is based on diagrammatic and covariant analyses which, in most cases, assume the so-called impulse approximation.
It has the advantage of producing GPDs covering the entire kinematic domain and fulfilling polynomiality, but is plagued with several issues, such as the lack of positivity or issues with discrete symmetries when dealing with momentum dependent vertex models (see Ref.~\cite{Mezrag:2016hnp} and references therein).
The second one is to use the expansion in Fock space in terms of Light-front wave-functions (LFWFs).
This way, positivity is naturally fulfilled as GPDs are given as an inner product of LFWFs, \emph{e.g.}~in the DGLAP region (see Refs.~\cite{Diehl:2000xz, Diehl:2003ny, Chouika:2017dhe,Chouika:thesis2018} for the notations and more details)
\begin{align}
  \label{eq:overlap-pion-quark-gpd-dglap}
  H^{q}\left(x,\xi,t\right) = &\sum_{N,\beta} \left(\sqrt{1-\xi^2}\right)^{2-N} \sum_a \delta_{f_a,q}
  \int\left[\drm \bar{x}\right]_{N}\left[\drm^{2}\vperp{\bar{k}}\right]_{N} \delta\left(x-\bar{x}_a\right) \\ \nonumber
  &\times \Psi^{*}_{N,\beta}\left(x^{\text{out} \prime}_{1},\vperp[1]{k}^{\text{out} \prime}, \dots,
  x^{\text{out} \prime}_{a},\vperp[a]{k}^{\text{out} \prime}, \dots\right)
  \Psi_{N,\beta}\left(x^{\text{in}}_{1},\vperp[1]{k}^{\text{in}}, \dots,
  x^{\text{in}}_{a},\vperp[a]{k}^{\text{in}}, \dots\right) \, ,
\end{align}
where LFWFs of the same number of partons overlap.
On the other hand, it is difficult to truncate in a consistent way in both the DGLAP ($\left|x\right| \ge \xi$) and ERBL ($\left|x\right| \le \xi$) regions.
Indeed, in the latter, the overlap is asymmetrical in the number $N$ of partons.
This renders polynomiality unlikely to be fulfilled, nor the GPD to be consistent in both regions (continuous at the border $\left|x\right| = \xi$ in particular), at any finite order of truncation $N$.
It is rather expected to be achieved when all Fock states are summed over.

\section{Covariant extension}
\label{sec:covext}

In practice, we often have a low order truncation, \emph{e.g.}~a two-body LFWF in the case of the pion.
Deriving the corresponding DGLAP GPD is straightforward.
Then, the problem can be stated as follows: what is the corresponding ERBL contribution? And how to reconstruct it?

For this, we can use the natural representation of the polynomiality property: Double Distributions (DDs).
Writing the GPD as a Radon transform \cite{Teryaev:2001qm,Radon:1917tr} in the following way~\cite{Pobylitsa:2002vi}:
\begin{equation}
\label{eq:GPD-DD-Pobylitsa}
H\left(x, \xi\right) = \left(1-x\right) \int_{\left|\alpha\right|+\left|\beta\right| \le 1} \mathrm{d}\beta \mathrm{d}\alpha \,
h(\beta, \alpha) \, \delta(x - \beta - \alpha \xi) \, ,
\end{equation}
we can use the DGLAP region to invert the equation and derive the DD $h$ which will allow us to extend the GPD then to the ERBL region~\cite{Chouika:2017dhe}.

 \begin{figure*}[t]
 \begin{tabular}{cc} 
  \includegraphics[width=0.45\textwidth]{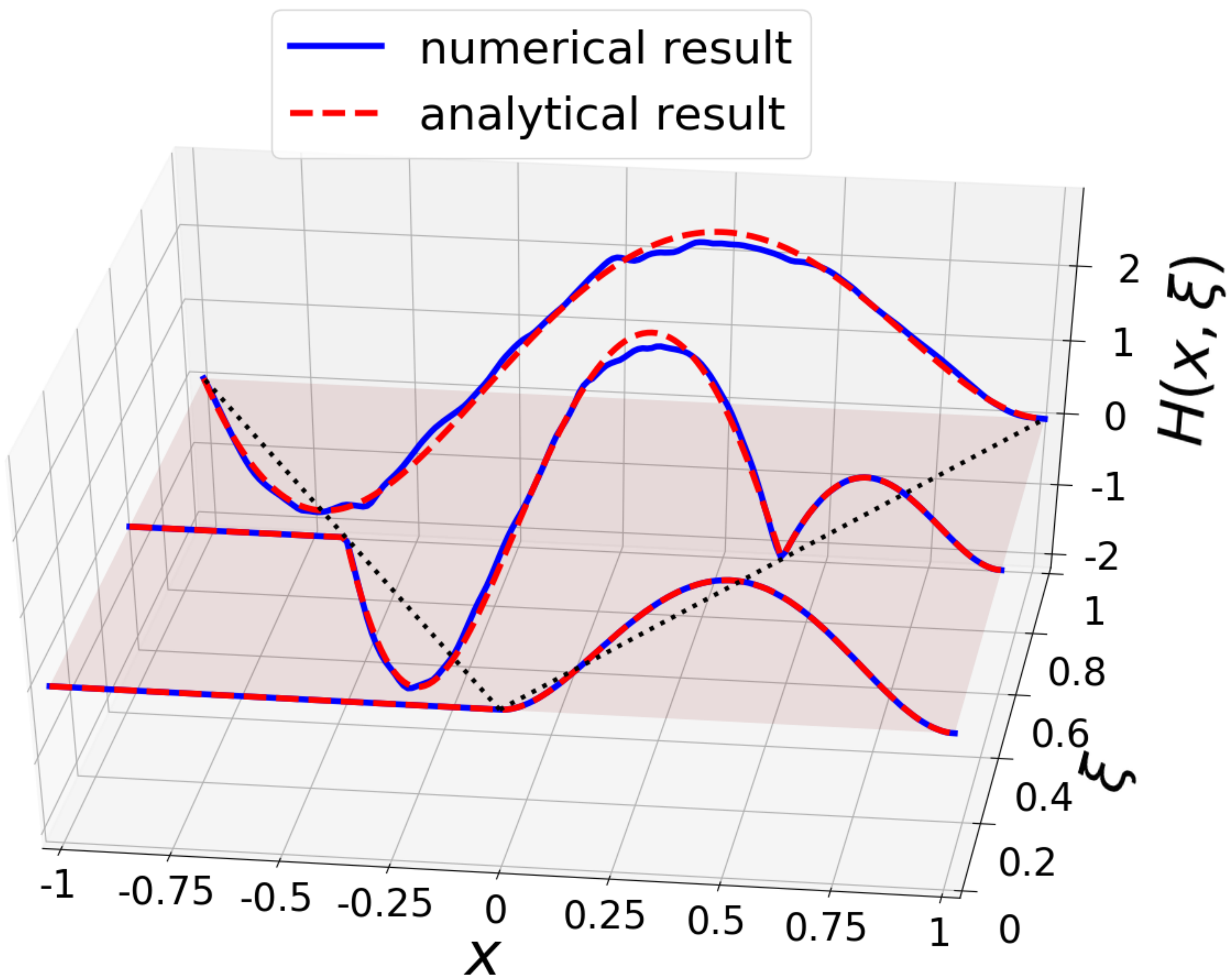} & 
  \includegraphics[width=0.45\textwidth]{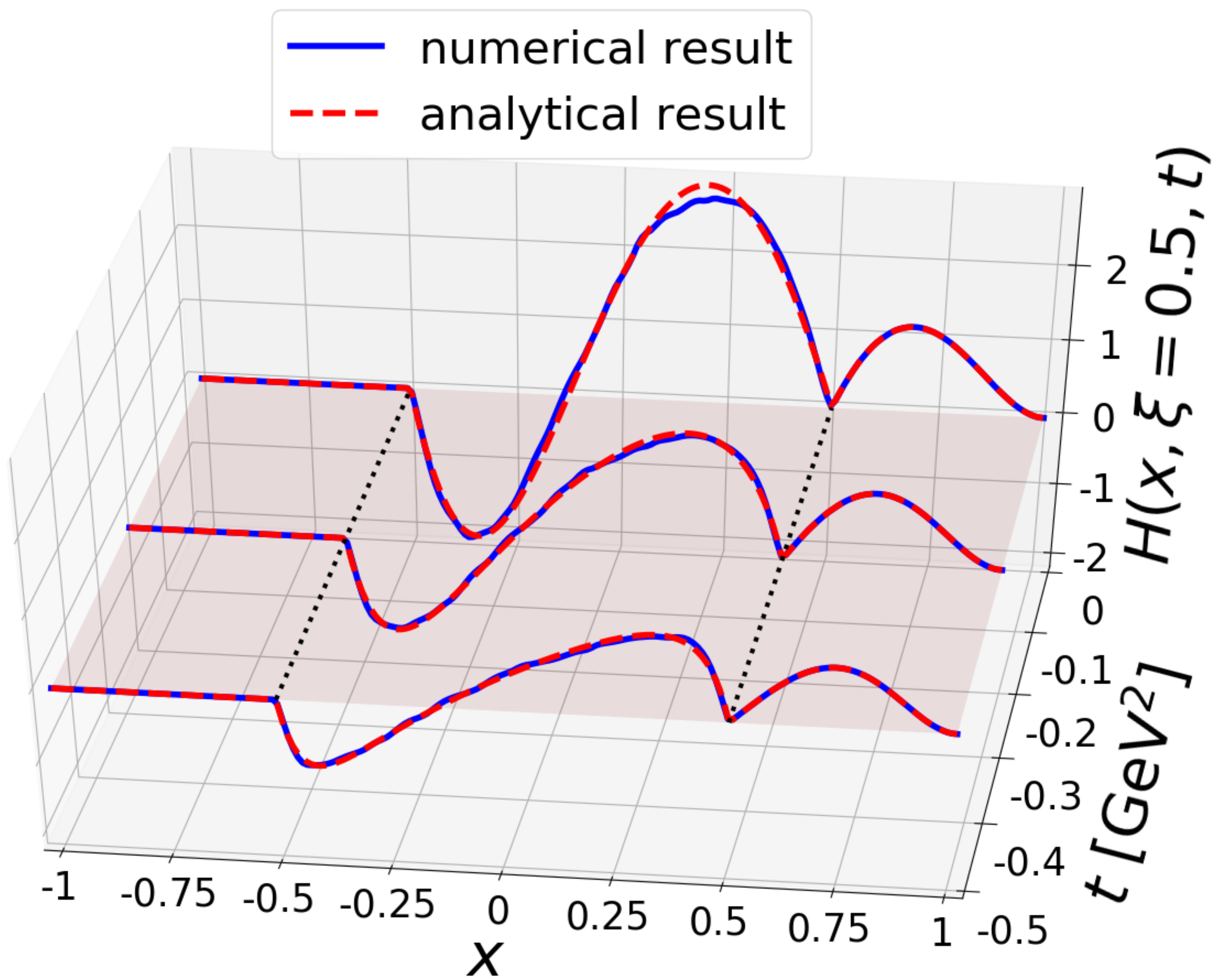}
 \end{tabular}
 \caption{
 Comparison between algebraic and numerical results for the pion GPD modeled in Refs.~\cite{Mezrag:2016hnp,Chouika:2017rzs}.
 The blue solid curves display the numerical results while the red dashed ones show the results algebraically derived in that case.
 The left panel stands for the case $t=0$ for fixed values of $\xi=\left[0,0.5,1\right]$ and the right one shows the $t$-behavior for fixed values $\left[0,-0.25,-0.5\right]$ at $\xi=0.5$.
 For more details, see Ref.~\cite{Chouika:2017dhe} where this figure was taken from.
 }
 \label{fig:cedric-algebraic-DSE-model}
 \end{figure*}

The main advantage of this method is that it is general in the sense that the inversion can be dealt with numerically, the procedure being the same for any input model of LFWFs.
Some models though allow for a simple algebraic guess of the DD and can therefore serve as benchmarks.
Fig.~\ref{fig:cedric-algebraic-DSE-model} illustrates this with a pion GPD built from a Nakanishi-based algebraic model.
The numerical inversion relies on the knowledge of the DGLAP region only, the extension to ERBL being our main goal.
Therefore, the examination of algebraic and numerical GPDs over the ERBL region (\emph{i.e.} between the black dotted lines) is the main outcome of this figure.
As can be seen, this numerical extrapolation is very good.
More details can be found in Refs.~\cite{Chouika:2017dhe,Chouika:thesis2018}.

\section{PARTONS framework}

\begin{wrapfigure}[9]{O}{0.35\columnwidth}
  \centering
  \includegraphics[width=0.34\columnwidth]{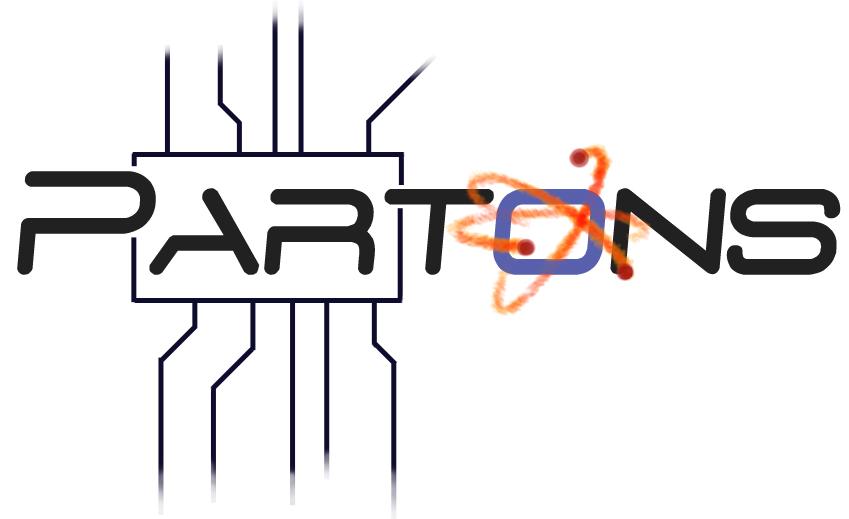}
  \caption{PARTONS logo. \label{fig:PARTONS-logo}}
\end{wrapfigure}

The PARTONS framework~\cite{Berthou:2015oaw} was conceived as an answer to the theoretical challenges facing GPD modelling in order to accompany the active experimental programs with foreseen increased accuracy and kinematic coverage.
It provides a C++ library aimed both at experimentalists and phenomenologists.

PARTONS encompasses the whole chain of computation of an observable in a given channel related to GPDs.
This can be divided into three main levels:
\begin{description}
  \item[Large distance] This level concerns the computation of GPDs themselves, with respect to different model parameters, as functions of $x$, $\xi$, $t$, etc.
  The factorization scale dependence is described by evolution equations.
  \item[Small distance] The second level is that of the small distance coefficient functions.
  In practice, it means convoluting the GPDs and the end results are the CFFs (see \emph{e.g.} Eq.~\eqref{eq:CFF-H-convolution}).
  \item[Full process] Finally, this level concerns the cross-sections and various other observables that can be directly accessed in experiments.
\end{description}
At any such level, the framework is flexible enough to allow any choice of model assumption, the inclusion of higher order corrections, etc.
Indeed, PARTONS is modular by design and works only on the basis of the needed abstract classes, unknowingly of what the user chooses for the physics content.

So far, only the DVCS channel is implemented in PARTONS, but the other exclusive processes (TCS and DVMP) are also planned, and the architecture was thought of to accommodate any such channel.
Among the available modules, we can cite the popular GK model for GPDs~\cite{Kroll:2012sm}, or the latest set of DVCS cross-section formulas in the Belitsky-Mueller formalism~\cite{Belitsky:2012ch}.
The current version of PARTONS has all the necessary tools to study DVCS at NLO and leading-twist, but other models and features can easily be added (or plugged by the user) due to its modularity.

\section{Phenomenology of quark models}

Now, making use of all the physics developments already implemented in PARTONS, we can build upon the covariant extension method presented in Sec.~\ref{sec:covext} to produce phenomenological outputs from constituent-quark-like models for instance, in the case of DVCS.
We choose here the Chiral Quark Soliton Model ($\chi$QSM) studied in Ref.~\cite{Lorce:2011dv}, truncated at the first Fock sector (three valence quarks).
This truncation implies that only the DGLAP region of GPDs is accessible, which limits the phenomenology to a leading-order (LO) analysis of DVCS, as only the cross-over line $x=\xi$ is needed in that case with the addition of (at least) a subtraction constant following the dispersion relations approach (see \emph{e.g.}~Ref.~\cite{Guidal:2013rya} and references therein).
We cannot go further in perturbation theory, nor evolve the GPD from the low scale $Q_0^2=0.259$~GeV$^2$ of the model to that of the experiment.

\begin{figure*}[t]
  \centering
  \begin{tabular}{@{}>{\centering}m{0.54\textwidth}@{}>{\centering}m{0.02\textwidth}@{}>{\centering}m{0.44\textwidth}@{}}
    \includegraphics[width=\linewidth]{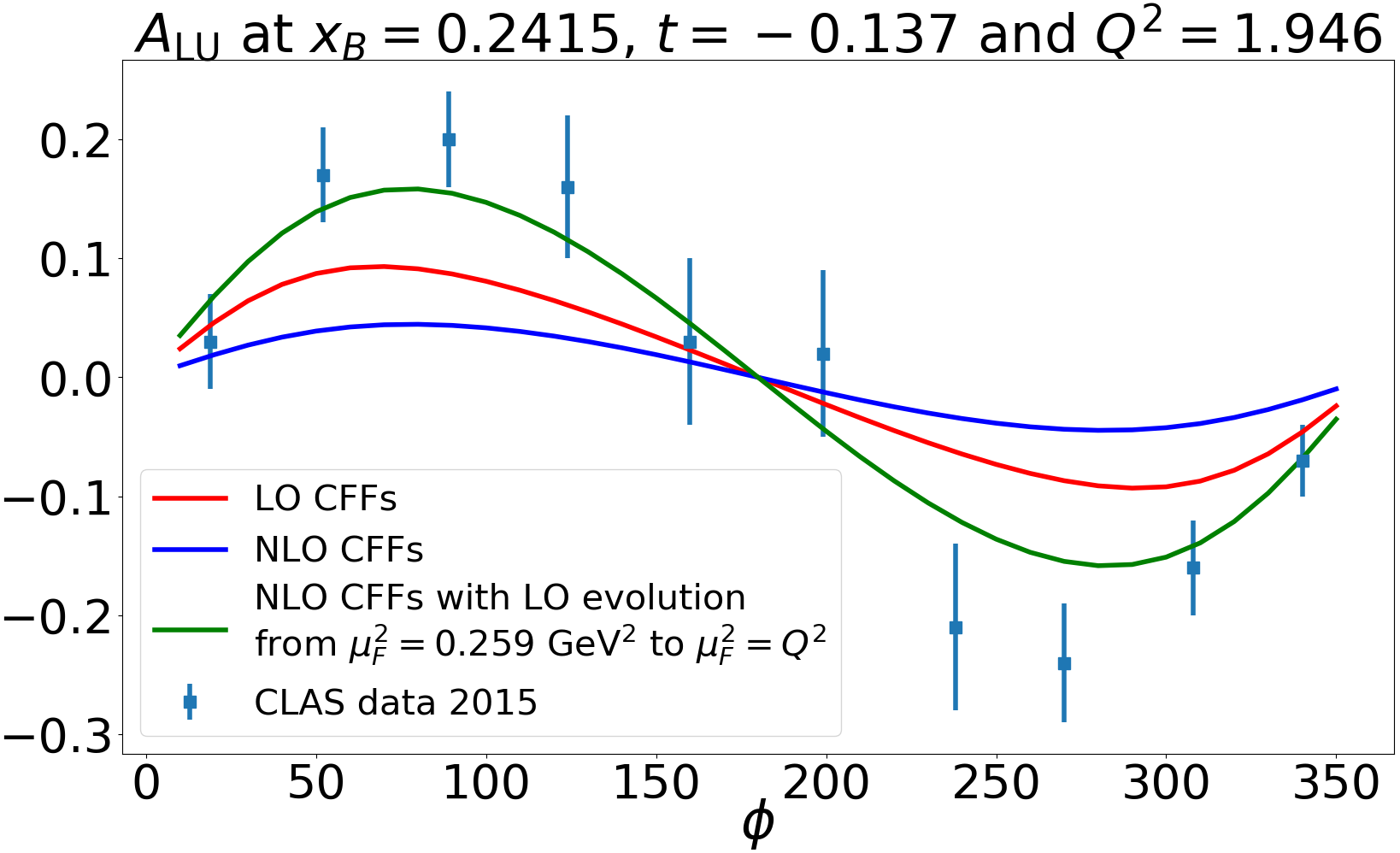}
    & &
    \includegraphics[width=\linewidth]{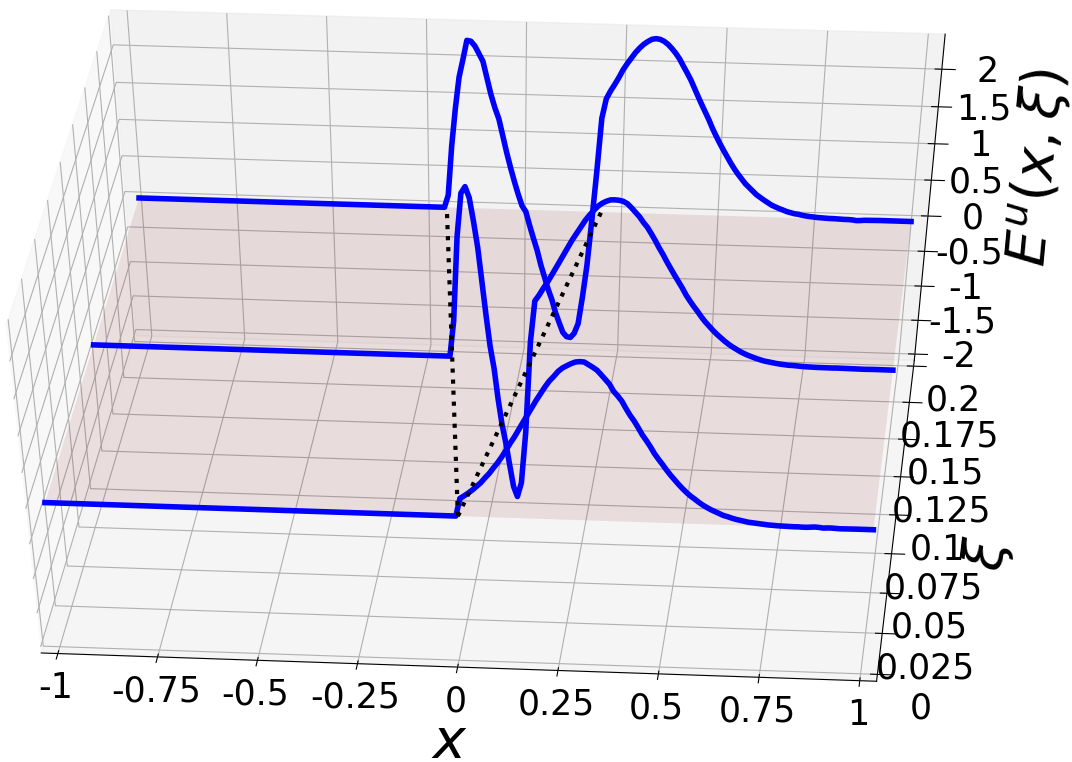}
  \end{tabular}
  \caption{Test of the Chiral Quark Soliton Model of Ref.~\cite{Lorce:2011dv}.
    \textbf{Left}: DVCS beam-spin asymmetry with CLAS data from Ref.~\cite{Pisano:2015iqa}.
    \textbf{Right}: GPD $E^u$ at $t=-0.34$~GeV$^2$ with extension to ERBL.
    \label{fig:plot-Alu-gpd-chqsm}}
\end{figure*}

By covariantly extending the GPD to the ERBL region, these studies become now possible.
We show in Fig.~\ref{fig:plot-Alu-gpd-chqsm} (right panel) this extension for the GPD $E$.
We can then use the covariantly extended GPDs $H$, $E$ and $\tilde{H}$\footnote{We neglect the GPD $\tilde{E}$.} for each flavor $u$ and $d$ to produce DVCS observables and compare them to experimental data.
In Fig.~\ref{fig:plot-Alu-gpd-chqsm} (left panel), we show as an example the calculation of the beam spin asymmetry defined in terms of the $e p \rightarrow e p \gamma$ cross-section as:
\begin{align}
  \label{eq:beam-spin-asymmetry}
  A_{\text{LU}}^{-}\left(\phi\right) = \frac{ \drm \sigma^{\overset{-}{\rightarrow}}\left(\phi\right)
                                                   - \drm \sigma^{\overset{-}{\leftarrow}}\left(\phi\right) }
                                                   { \drm \sigma^{\overset{-}{\rightarrow}}\left(\phi\right)
                                                   + \drm \sigma^{\overset{-}{\leftarrow}}\left(\phi\right) } \, ,
\end{align}
where $\phi$ is the angle between the lepton plane and the production plane, the arrows denote the helicity and the minus sign the charge of the beam\footnote{At JLab, only an electron beam is available.}.
The JLab kinematics chosen are part of a set of data published by the CLAS collaboration~\cite{Pisano:2015iqa}.

We do not say anything about the accuracy of this specific model here, as our present goal is only to illustrate this procedure.
It can be applied systematically to any given valence quark models to produce GPDs fulfilling all theoretical constraints and ready for DVCS phenomenology under various perturbative hypotheses.
As we expect in the near future the publication of the new JLab 12 GeV data for the valence region, this study would be most welcome.
We could indeed test the relevance at low scale of a valence truncation of LFWFs and pave the way for a systematic phenomenology of LFWFs through exclusive processes and GPDs.
This could potentially lead \emph{in fine} to a unified phenomenology of both GPDs and TMDs.

\bibliographystyle{JHEP}
\bibliography{biblio}

\providecommand{\href}[2]{#2}\begingroup\raggedright\begin{thebibliography}{10}

\bibitem{Mueller:1998fv}
D.~Mueller, D.~Robaschik, B.~Geyer, F.~Dittes and J.~Ho\v{r}e\v{j}si\emph{},
  \href{https://doi.org/10.1002/prop.2190420202}{\emph{Fortsch.Phys.}
  {\bfseries 42} (1994) 101}
  [\href{https://arxiv.org/abs/hep-ph/9812448}{{\ttfamily hep-ph/9812448}}].

\bibitem{Radyushkin:1996nd}
A.~Radyushkin\emph{},
  \href{https://doi.org/10.1016/0370-2693(96)00528-X}{\emph{Phys.Lett.}
  {\bfseries B380} (1996) 417}
  [\href{https://arxiv.org/abs/hep-ph/9604317}{{\ttfamily hep-ph/9604317}}].

\bibitem{Ji:1996nm}
X.-D. Ji\emph{},
  \href{https://doi.org/10.1103/PhysRevD.55.7114}{\emph{Phys.Rev.} {\bfseries
  D55} (1997) 7114} [\href{https://arxiv.org/abs/hep-ph/9609381}{{\ttfamily
  hep-ph/9609381}}].

\bibitem{Burkardt:2000za}
M.~Burkardt\emph{}, \href{https://doi.org/10.1103/PhysRevD.62.071503,
  10.1103/PhysRevD.66.119903}{\emph{Phys. Rev.} {\bfseries D62} (2000) 071503}
  [\href{https://arxiv.org/abs/hep-ph/0005108}{{\ttfamily hep-ph/0005108}}].

\bibitem{Belitsky:2001ns}
A.~V. Belitsky, D.~Mueller and A.~Kirchner\emph{},
  \href{https://doi.org/10.1016/S0550-3213(02)00144-X}{\emph{Nucl.Phys.}
  {\bfseries B629} (2002) 323}
  [\href{https://arxiv.org/abs/hep-ph/0112108}{{\ttfamily hep-ph/0112108}}].

\bibitem{Diehl:2003ny}
M.~Diehl\emph{},
  \href{https://doi.org/10.1016/j.physrep.2003.08.002}{\emph{Phys.Rept.}
  {\bfseries 388} (2003) 41}
  [\href{https://arxiv.org/abs/hep-ph/0307382}{{\ttfamily hep-ph/0307382}}].

\bibitem{Belitsky:2005qn}
A.~Belitsky and A.~Radyushkin\emph{},
  \href{https://doi.org/10.1016/j.physrep.2005.06.002}{\emph{Phys.Rept.}
  {\bfseries 418} (2005) 1}
  [\href{https://arxiv.org/abs/hep-ph/0504030}{{\ttfamily hep-ph/0504030}}].

\bibitem{Pire:1998nw}
B.~Pire, J.~Soffer and O.~Teryaev\emph{},
  \href{https://doi.org/10.1007/s100529901063}{\emph{Eur.Phys.J.} {\bfseries
  C8} (1999) 103} [\href{https://arxiv.org/abs/hep-ph/9804284}{{\ttfamily
  hep-ph/9804284}}].

\bibitem{Mezrag:2016hnp}
C.~Mezrag, H.~Moutarde and J.~Rodriguez-Quintero\emph{},
  \href{https://doi.org/10.1007/s00601-016-1119-8}{\emph{Few Body Syst.}
  {\bfseries 57} (2016) 729}
  [\href{https://arxiv.org/abs/1602.07722}{{\ttfamily 1602.07722}}].

\bibitem{Diehl:2000xz}
M.~Diehl, T.~Feldmann, R.~Jakob and P.~Kroll\emph{},
  \href{https://doi.org/10.1016/S0550-3213(00)00684-2}{\emph{Nucl.Phys.}
  {\bfseries B596} (2001) 33}
  [\href{https://arxiv.org/abs/hep-ph/0009255}{{\ttfamily hep-ph/0009255}}].

\bibitem{Chouika:2017dhe}
N.~Chouika, C.~Mezrag, H.~Moutarde and J.~Rodríguez-Quintero\emph{},
  \href{https://doi.org/10.1140/epjc/s10052-017-5465-6}{\emph{Eur. Phys. J.}
  {\bfseries C77} (2017) 906}
  [\href{https://arxiv.org/abs/1711.05108}{{\ttfamily 1711.05108}}].

\bibitem{Chouika:thesis2018}
N.~Chouika, \emph{Generalized Parton Distributions and their covariant
  extension: towards nucleon tomography}, Ph.D. thesis, Université
  Paris-Saclay, 2018.

\bibitem{Teryaev:2001qm}
O.~Teryaev\emph{},
  \href{https://doi.org/10.1016/S0370-2693(01)00564-0}{\emph{Phys.Lett.}
  {\bfseries B510} (2001) 125}
  [\href{https://arxiv.org/abs/hep-ph/0102303}{{\ttfamily hep-ph/0102303}}].

\bibitem{Radon:1917tr}
J.~Radon\emph{}, \href{https://doi.org/10.1109/TMI.1986.4307775}{\emph{Medical
  Imaging, IEEE Transactions on} {\bfseries 5} (1986) 170}.

\bibitem{Pobylitsa:2002vi}
P.~Pobylitsa\emph{},
  \href{https://doi.org/10.1103/PhysRevD.67.034009}{\emph{Phys.Rev.} {\bfseries
  D67} (2003) 034009} [\href{https://arxiv.org/abs/hep-ph/0210150}{{\ttfamily
  hep-ph/0210150}}].

\bibitem{Chouika:2017rzs}
N.~Chouika, C.~Mezrag, H.~Moutarde and J.~Rodríguez-Quintero\emph{},
  \href{https://doi.org/10.1016/j.physletb.2018.02.070}{\emph{Phys. Lett.}
  {\bfseries B780} (2018) 287}
  [\href{https://arxiv.org/abs/1711.11548}{{\ttfamily 1711.11548}}].

\bibitem{Berthou:2015oaw}
B.~Berthou et~al.\emph{},
  \href{https://doi.org/10.1140/epjc/s10052-018-5948-0}{\emph{Eur. Phys. J.}
  {\bfseries C78} (2018) 478}
  [\href{https://arxiv.org/abs/1512.06174}{{\ttfamily 1512.06174}}].

\bibitem{Kroll:2012sm}
P.~Kroll, H.~Moutarde and F.~Sabatie\emph{},
  \href{https://doi.org/10.1140/epjc/s10052-013-2278-0}{\emph{Eur.Phys.J.}
  {\bfseries C73} (2013) 2278}
  [\href{https://arxiv.org/abs/1210.6975}{{\ttfamily 1210.6975}}].

\bibitem{Belitsky:2012ch}
A.~V. Belitsky, D.~Müller and Y.~Ji\emph{},
  \href{https://doi.org/10.1016/j.nuclphysb.2013.11.014}{\emph{Nucl.Phys.}
  {\bfseries B878} (2014) 214}
  [\href{https://arxiv.org/abs/1212.6674}{{\ttfamily 1212.6674}}].

\bibitem{Lorce:2011dv}
C.~Lorcé, B.~Pasquini and M.~Vanderhaeghen\emph{},
  \href{https://doi.org/10.1007/JHEP05(2011)041}{\emph{JHEP} {\bfseries 1105}
  (2011) 041} [\href{https://arxiv.org/abs/1102.4704}{{\ttfamily 1102.4704}}].

\bibitem{Guidal:2013rya}
M.~Guidal, H.~Moutarde and M.~Vanderhaeghen\emph{},
  \href{https://doi.org/10.1088/0034-4885/76/6/066202}{\emph{Rept.Prog.Phys.}
  {\bfseries 76} (2013) 066202}
  [\href{https://arxiv.org/abs/1303.6600}{{\ttfamily 1303.6600}}].

\bibitem{Pisano:2015iqa}
{\scshape CLAS} collaboration, S.~Pisano et~al.\emph{},
  \href{https://doi.org/10.1103/PhysRevD.91.052014}{\emph{Phys. Rev.}
  {\bfseries D91} (2015) 052014}
  [\href{https://arxiv.org/abs/1501.07052}{{\ttfamily 1501.07052}}].

\end{thebibliography}\endgroup

\end{document}